\documentclass[
    reprint, 
    amsmath,
    amssymb, 
    aps, 
    twoside, 
    superscriptaddress,
]{revtex4-1}

\usepackage[T1]{fontenc} 
\usepackage{physics}
\usepackage{dsfont} 
\usepackage{color,newfloat}
\usepackage{graphicx}
\usepackage{dcolumn}
\usepackage{bm}
\usepackage{bbm}
\usepackage{amsmath,amssymb}
\usepackage{subfiles}

\usepackage[nonumberlist, acronym, toc]{glossaries}
\usepackage[inline]{enumitem}
\usepackage[ampersand]{easylist}

\usepackage{array}  
\usepackage{hhline}
\usepackage{multirow}
\usepackage{afterpage}
\usepackage[normalem]{ulem}
\usepackage{float}
\usepackage{nomencl}
\makenomenclature
\usepackage{etoolbox}

\usepackage[UKenglish]{babel}

\usepackage{todonotes}
\usepackage{soul}

\setcitestyle{super}

\usepackage{siunitx}
\usepackage[ruled]{algorithm2e} 
\usepackage{newfloat,algcompatible} 
\usepackage{xcolor}

\usepackage[pdftex,colorlinks=true,bookmarks=false,citecolor=blue,urlcolor=blue]{hyperref}

\def\({\left(}
\def\){\right)}
\def\[{\left[}
\def\]{\right]}  

\usepackage{titlesec}
\titleformat{\subsection}[runin]{\bfseries}{}{0pt}{}[]

\usepackage{mathtools}

\begin{document}

\title{
Robotic vectorial field alignment for spin-based quantum sensors.
}

\preprint{APS/123-QED}
\author{Joe A. Smith}
\email{j.smith@bristol.ac.uk}
\affiliation{Quantum Engineering Technology Labs and Department of Electrical and Electronic Engineering, University of Bristol, Bristol, BS8 1FD, UK}
\author{Dandan Zhang}
\affiliation{Bristol Robotics Laboratory and Department of Engineering Mathematics, University of Bristol, Bristol,  BS8 1TW, UK}
\author{Krishna C. Balram}
\email{krishna.coimbatorebalram@bristol.ac.uk}
\affiliation{Quantum Engineering Technology Labs and Department of Electrical and Electronic Engineering, University of Bristol, Bristol, BS8 1FD, UK}

\date{\today}


\begin{abstract}
Developing practical quantum technologies will require the exquisite manipulation of fragile systems in a robust and repeatable way. As quantum technologies move towards real world applications, from biological sensing to communication in space, increasing experimental complexity introduces constraints that can be alleviated by the introduction of new technologies. 
Robotics has shown tremendous progress in realising increasingly smart, autonomous and highly dexterous machines. Here, we demonstrate that a robotic arm equipped with a magnet can sensitise an NV centre quantum magnetometer in challenging conditions unachievable with standard techniques. We generate vector magnetic field with $1^\circ$ angular and 0.1 mT amplitude accuracy and determine the orientation of a single stochastically-aligned spin-based sensor in a constrained physical environment. Our work opens up the prospect of integrating robotics across many quantum degrees of freedom in constrained settings, allowing for increased prototyping speed, control, and robustness in quantum technology applications. 
\end{abstract}

\maketitle




Experiments designed to exploit quantum technologies for applications can be extremely challenging. Fragile quantum states must be delicately manipulated, whilst minimising sources of decoherence, in order to preserve a quantum advantage. This often necessitates cutting-edge experimental physics, including precise and complex optical assemblies \cite{qiang2018large,zhong2021phase}, strong vector magnetic fields \cite{soshenko2021nuclear}, high-speed microwave delivery \cite{kielpinski2002architecture}, and compatibility with extremely low temperature environments \cite{krinner2019engineering}. Emerging quantum technologies based on hybrid quantum systems \cite{kurizki2015quantum}  combine research from two or more experimental settings: such as coupling spins in silicon to superconducting resonators and qubits \cite{clerk2020hybrid, albertinale2021detecting}, interfacing remote NV centres in diamond with photonic qubits \cite{hermans2022qubit}, and using nanomechanics to interface with spins \cite{whiteley2019spin} or superconducting qubits \cite{balram2022piezoelectric}.

As these proof-of-principle devices become more sophisticated and start to scale in size and complexity, established lab infrastructure such as translation stages and solenoid coils will no longer provide the flexibility, speed, and precision to meet these constrained \cite{akhtar2023high} and sometimes competing experimental requirements. In contrast, the field of robotics has long adapted to operate robots in challenging conditions, such as at the microscale \cite{zhang2022micro} or in very low temperature environments \cite{cline2022lando}. Robotics can provide more flexible and adaptable approaches than traditional methods, that would speed up the deployment of quantum technology across applications.  With sophisticated software stacks and well-developed open-source hardware, the deployment of robotics in a diverse range of experimental settings in the chemical and biological sciences has become increasingly feasible \cite{granda2018controlling,durrer2022robot}. 

\begin{figure*}
\centering
\includegraphics[width=\textwidth]{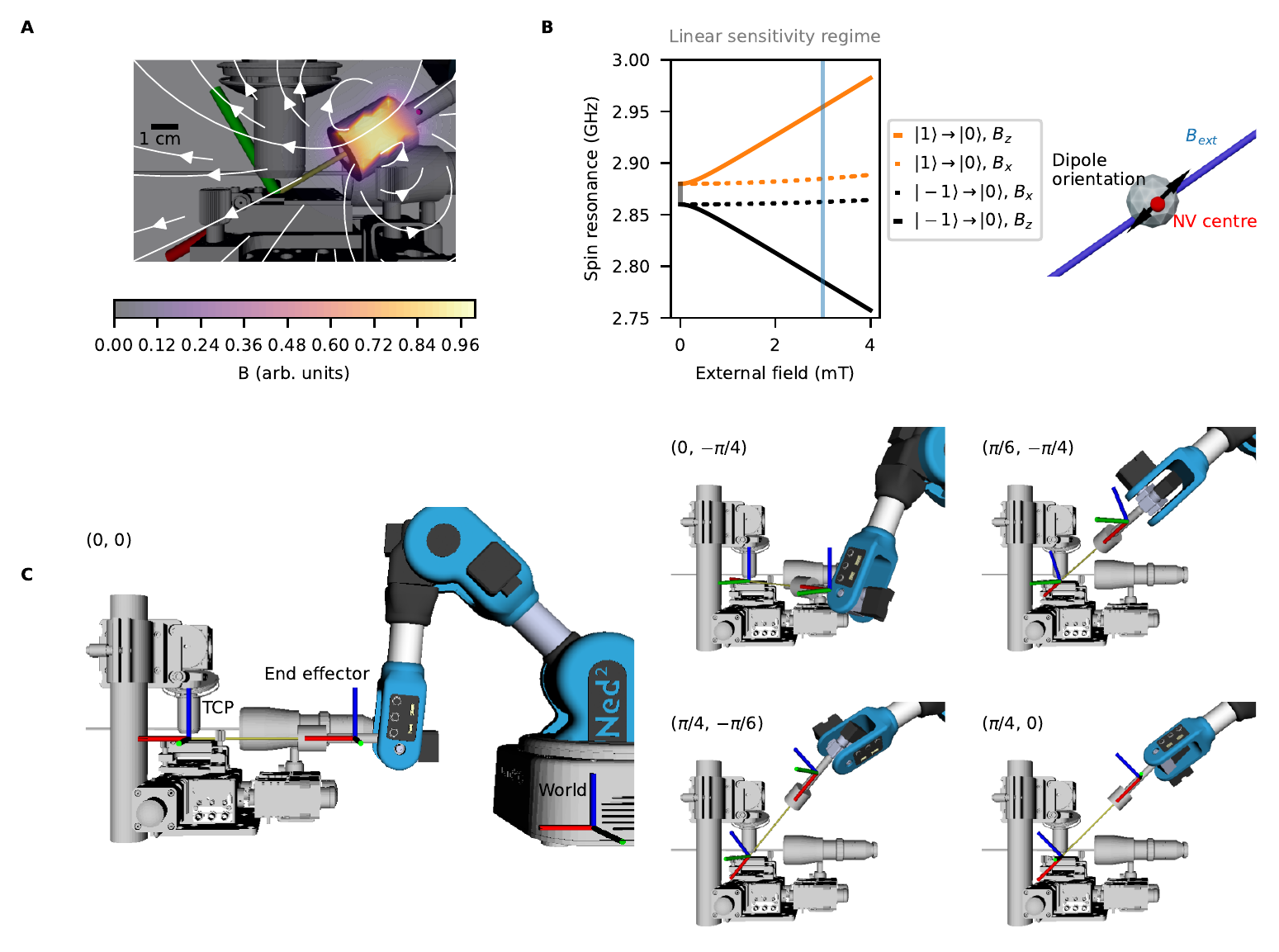}
\caption{\label{fig:setup} \textbf{Experimental setup and working principles.} \textbf{A.} Placing a permanent magnet near the NV centre magnetometer produces a magnetic field of a known orientation, defined along its axis (field lines in white). \textbf{B.}
One use case here is to change the spin resonance of the magnetometer, to operate at its most sensitive regime (linear with respect to detected field) away from the zero-field splitting (marked in grey). As observed, field along the NV centre $B_z$ affects this response, whereas transverse components only contribute unwanted performance degradation. The field $B_{ext}$ should therefore be approximately aligned to the NV centre magnetic dipole orientation $B_z$. 
\textbf{C.} The 6 DoF robot is used to orient the magnet in complex surroundings. The robot base is located at the world origin (x-axis indicated in red,  y-axis indicated in green, z-axis in blue). The Tool Centre Point (TCP axis marked) is translated along the x-axis of the end effector axis (marked) to set the required field strength. The TCP coordinate $(x, y, z, \alpha_y, \alpha_z)$ is then set to the NV centre position and rotated around the $y$ and $z$ axis, i.e. varying $\alpha_y$ and $\alpha_z$ of the TCP, to form a defined vector from the end effector to the TCP (shown in yellow). 
The robot position at a range of different $(\alpha_y,\alpha_z)$ are shown in the inset diagrams. Through this method the highly-dexterous robot can create fields with arbitrary field strengths and orientations, and align the TCP axis with the NV axis to produce the desired $B_z$. 
}
\end{figure*}

Here, we introduce and validate the idea of a robot-assisted quantum technology. Specifically, we employ the use of a robotic arm to hold a strong permanent magnet for meeting a requirement in spin-based sensing: aligning an external magnetic field along the magnetic dipole axis of an arbitrarily oriented spin system (Fig. \ref{fig:setup}A). We demonstrate that this method has significant advantages where traditional techniques for generating vector fields, such as mounting the magnet on a fixed axis translation stage, or using a 3-axis Helmholtz coil, are infeasible owing to the tight physical constraints of the surrounding optomechanical apparatus. While this work focuses on a specific use case for robotics in quantum technology, the methods developed here can be easily adapted and extended to other experimental settings.

\noindent\textbf{Problem statement and requirements}\\
Spin-based magnetometers operate by mapping local perturbations in their environment to shifts in the transition (magnetic resonance absorption) frequency of the spin system \cite{degen2017quantum}. The NV centre in diamond is the prototypical solid state quantum sensor on account of its optically accessible spin state which allows optical manipulation and readout of its spin state at room temperature (Optically Detected Magnetic Resonance or ODMR). NV centre magnetometers have rapidly advanced over the past decade and have reached maturity as a quantum magnetometer, with nanotesla (nT) sensitivities at nanoscale resolutions \cite{rondin2014magnetometry,casola2018probing}. As magnetic dipole-dipole interactions are weak and confined to the near field, near-surface NV centres are required to image fields from individual spins \cite{ajoy2015atomic}. Nanoscale inclusions of diamond, or nanodiamond, are used to host the spin-probe in hot and wet biochemical surroundings in applications such as protein \cite{ermakova2013detection,lovchinsky2016nuclear} or cell detection \cite{
mcguinness2011quantum,nie2021quantum}. 
Nanodiamonds typically contribute an additional energy term $\Pi$ to the NV centre Hamiltonian $H$, from lattice strain and local charges  \cite{alkahtani2019growth,mittiga2018imaging}:
\begin{align}\label{eq:Hamil}
{H}=DS_z^2+\Pi\left(S_x^2-S_y^2\right)+\gamma \mathbf{B}_{\perp} \cdot \mathbf{S}_{\perp}+ \gamma B_z S_z,
\end{align}
where $\mathbf{B}_{\perp}=\left(B_x, B_y\right)$ and $\mathbf{S}_{\perp}=\left(S_x, S_y\right)$ are the transverse magnetic field and Pauli spin terms, with $z$ defined as the axis comprising the NV centre along the diamond lattice.
In Fig. \ref{fig:setup}B, the energy term $\Pi$ leads to a frequency splitting of size $2\Pi$ (shown in grey), making the NV centre transition frequencies robust to magnetic field fluctuations to first order. A bias field $B_z$ is thus required bring to the NV centre into the regime ($B_z\gg{\Pi/\gamma}$) where the transitions are linearly dependent on the magnetic field, which corresponds to the highest sensitivity. Given nanodiamonds typically display $\Pi\sim 10$ MHz \cite{dolde2011electric}, this requires a moderate $B_z$ magnetic field of 5 mT aligned to the NV axis. A misaligned magnetic field (with a residual $B_x$ or $B_y$ component) would lead to a mixing of the energy eigenstates, which would result in a reduction of both the fluorescence and contrast (SNR) of the spin-dependent optical readout \cite{tetienne2012magnetic}. Magnetic fields of 5 mT significantly degrade the spin coherence time ($T_2$) of the NV centre when misaligned by 5$^\circ$ as they cause nearby nuclear spins to precess \cite{stanwix2010coherence,maze2008electron}.

To date, the established method to align a static magnetic field to an arbitrarily oriented spin is using three perpendicular wire coils \cite{geiselmann2013three,schloss2018simultaneous,fukushige2020identification} or sets of coils in the Helmholtz configuration \cite{stefan2021multiangle,knauer2020situ,dolde2011electric}. 
To produce appreciable field strengths ($\approx 10$ mT), coils comprise of hundreds of wire turns. The wire gauge is chosen to balance wire packing and current density. Coils operate hot, which has adverse implications for sensitive samples, such as in biosensing\cite{petrini2020quantum}. The configuration is convenient for producing vector magnetic fields, after calibration, as the current in each coil can be ratioed to produce a desired orientation. The constraint of requiring coils at three axes around the sample severely restricts optical or mechanical degrees of freedom. An alternative method is to place and orient a strong neodymium permanent magnet (NdFeB) in the vicinity of the sample. The advantage here is that the small magnet can produce much larger field strengths than the coil. It is less restrictive in physical footprint so can be combined with optical assemblies and cryogenics. The magnets are aligned using linear \cite{weggler2020determination,stark2017development,cooper2020identification} or rotational translation stages \cite{holzgrafe2020nanoscale,wood2022long}. The physical limitations of these stages preclude certain orientation NV centres from being aligned \cite{wood2022long}. Typically, the magnet is positioned once and is aligned along a set diamond crystalline axis because the calibration process is cumbersome. This precludes the use of nanodiamonds where each site may have a random dipole orientation \cite{aslam2023quantum}, eliminating important applications that involve inspecting spatially separated regions or tracking dynamic events in liquid environments. 

We propose to combine the convenience and control of coils with the small footprint and strength of a permanent magnet in developing a robotically controlled vectorial field alignment system. Our approach has the following advantages:
(1) \textbf{Increased precision and control}. The robot manipulates the magnet with a high degree of accuracy, ensuring precise alignment of the generated magnetic field. For our application, this means better than $5 ^\circ$ accuracy \cite{stanwix2010coherence}. (2) \textbf{Fast alignment}: by employing a robot to move and position the magnet across optimal trajectories, alignment should be more efficient than manual techniques. (3) \textbf{Long-term stability}: employing closed-loop feedback with sufficient torque against gravity will maintain position securely for extended periods, ensuring stable alignment during experiments. (4) \textbf{Enhanced reproducibility}: A robust algorithm can align and realign the magnet between sample exchange or across multiple sites of interrogation. The robot consistently produces a given orientation field for different sample geometries. (5) \textbf{Scalability}: the robot has an adaptable routine that is able to suit a wide range of experimental configurations and constraints. One can also easily extend to scenarios where two fields with specific orientations need to be simultaneously applied, for instance an in-plane and out-of-plane field.

For positioning of a magnet at a desired location in 3D space with respect to a point of interest, and allowing for rotation about two axes to achieve magnetic field orientation, we require a robot with at least five degrees of freedom. The robot must be capable of handling a moderate payload in order to carry enough magnet mass to produce an appreciable field (10 mT) at a distance. It should also be readily available, with a well developed software interface and be economical to meet the requirements of use outside the robotics community. In the following section we will evaluate this robot for the described task.  

\section{Results}

\begin{figure*}
\centering
\includegraphics[width=\textwidth]{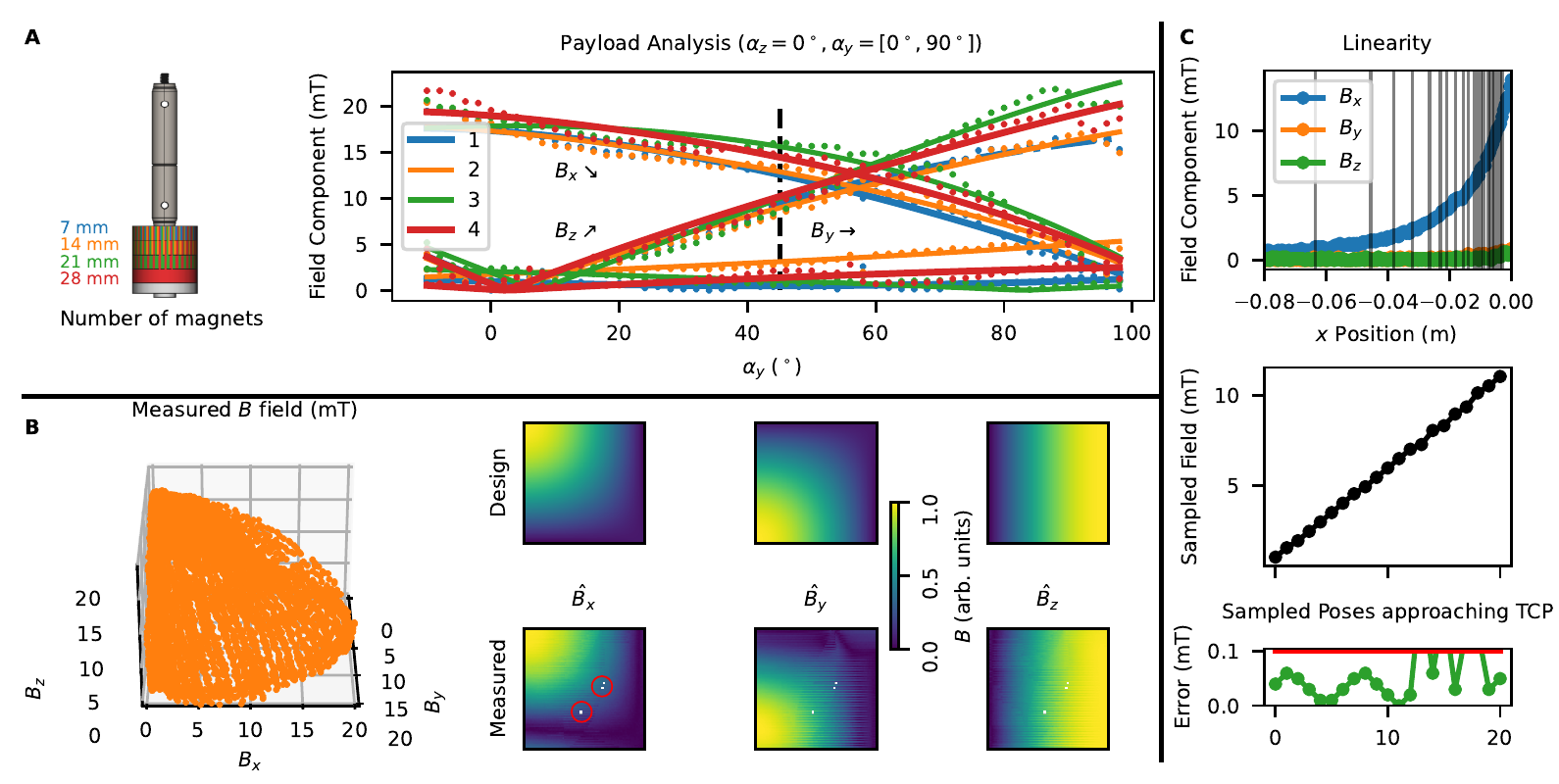}
\caption{\label{fig:hall} \textbf{Robot arm generates arbitrary vector magnetic fields.} 
\textbf{A.} A permanent magnet of varying mass is placed in the tool (left panel). A Hall sensor measures the $x-z$ trajectory of the field produced by the arm (right panel). The trajectory is well-fitted using a model of the field generated by the cylindrical magnet, noting a 15 $^\circ$ offset in the $x$-$z$ crossing from the expect 45$^\circ$ (shown with dotted line) and a varying non-zero offset in $y$, with this offset resulting in a non-linear trend in field registered with increasing magnetic mass. This initial measurement and model can be used for fine alignment calibration.  
\textbf{B.} The arm can create a field over the full $x-y-z$ sphere segment (one-eighth) with 3$^\circ$ accuracy. White pixels in the image plots (circled in red) indicate the few unreachable positions. \textbf{C.} The distance $r$ from the end effector to the Tool Center Point (TCP) produces a field strength fall off in $B_x$ proportional to $1/r^3$ (top panel), from which points (shown by vertical lines) can be then sampled (middle panel) to create a linear field response with high accuracy (bottom panel). }
\end{figure*}

\subsection{Workspace analysis}
Workspace analysis is essential for robot control and application, as it evaluates the space the robot can access and manipulate with its end-effector, constrained by the robot’s kinematic configuration. This analysis helps to identify the robot's suitability for specific tasks and environments. Key aspects include the reachable workspace volume (total 3D space the robot's end-effector can reach), workspace boundaries (limits of reachable space) and dexterity within this workspace (ability to precisely orient the end-effector) \cite{zhang2019wsrender}. In this section, we perform workspace analysis on a magnet carrying robotic arm to evaluate its performance in generating vector magnetic fields.

The robotic arm consists of a set of rigid bodies called links, connected by joints, with each joint driven by a motor actuator. An end-effector, in this case a permanent magnet, is attached to the end link. The arm is an open chain robot, with the position and orientation of the end-effector uniquely determined from the joint positions. The common configuration comprises of six joints, providing six Degrees of Freedom (DoF). For our experiments, we use a \emph{Niryo NED 2} robot owing to its well-documented open source stack, and ready availability. The arm has a moderate payload of 300 g, and is thus capable of lifting 40 cm$^3$ NdFeB, which can generate a surface magnetic field of $\approx$ 800 mT (see Supplementary Fig. 1). For ease of adoption, we select a cylindrical magnetic source with a radial hole, through which it can be fixed by a screw to the tool shaft. Shown in Fig. \ref{fig:setup}C, the robot is first set up by translating the Tool Center Point (TCP) along the x-axis of its end-effector, coaxial with the magnetisation axis of the magnet (Fig. \ref{fig:setup}A). This translation sets the distance, and hence strength, of the magnet, from the point of interrogation. 

To create a set vector field, the robot uses the Robot Operating System (ROS) kinematic processor \cite{quigley2009ros} to position its joints, in order to compute a desired pose. The robot pose comprises the location and orientation of the TCP relative to a global coordinate frame. Rigid robots possess six state variables $(x, y, z, \alpha_x, \alpha_y, \alpha_z)$, where the latter three coordinates are angles of rotation about the x, y and z axis respectively. The inverse kinematics problem is to find the joint position given a desired pose. In Supplementary Table 1, we give the Denavit-Hartenberg (D-H) representation for the kinematics of this robot. In principle, by fixing $x$, $y$, and $z$ at the NV centre location, the vector orientation of an applied magnetic field can be modified by varying $\alpha_y$ and $\alpha_z$ of the pose. In our scheme, the cylindrical magnet is symmetric about $\alpha_x$, so this degree of freedom is left unused. In Fig. \ref{fig:setup}C, we simulate in RViz, a visualization software for ROS, that the robot is sufficiently dexterous in positioning its joints to achieve a range of orientations, whereby the magnet is rotated around a stationary point with varying $\alpha_y$ and $\alpha_z$. In Supplementary Fig. 2, we calculate the full workspace volume and dexterity within this volume.  

\subsection{Magnetic vector reconstruction}
The goal in controlling the pose angle is to create a desired magnetic vector field at a given sample location. To experimentally verify this, we position a 3-axis Hall sensor at the point of interest in order to measure the field generated by the robot. We set the robot approximately collinear with the sensor axis, observed using a camera with a zoom lens. In Fig. \ref{fig:hall}A, we see the effect of adding magnets to the structure up to  70 \% of the payload by setting the robot along an arc trajectory from horizontal to vertical, by rotating the desired pose from $\alpha_y = 0$ to $\alpha_y = \pi/2$ with the distance between the sample and magnet surface fixed.

Commonly in robotics, camera data is processed to extract information on the desired pose \cite{zhang2022micro}. Here, the 3-axis Hall Sensor provides rich additional vector information, which, coupled with the known dependence of the magnetic field on position, allows the desired pose to be measured with higher precision than is visually observable. We fit the data with a closed form expression of the magnetic field observed from the cylindrical magnet \cite{magpylib2020}, using the pose variables as fitting parameters. We fit a constant 15$^\circ$ offset in $\alpha_y$ and observe this in the $x-z$ crossing point of the sensor and the robot, which for an aligned system would occur at 45$^\circ$ (marked by a dashed line in Fig. \ref{fig:hall}A). Additionally, for this trajectory, we would expect no $B_y$ field to be measured. The non-zero $B_y$ component is well fit to a varying non-zero $\alpha_z$ occurring when each magnet is added. This offset results in a non-linear relation between the number of magnets added and the observed strength. With an initial calibration trajectory to record this magnetic field information, fine alignment can achieved either by physically adjusting the robot or by modifying the coordinate frame to correct for the observed misalignment error.  

Following this initial trajectory, in Fig. \ref{fig:hall}B we observe that by scanning through a dictionary of poses, varying only $\alpha_y$ and $\alpha_z$, we are able to traverse a set of $B_{abs} \hat{\mathbf{n}}(\alpha_y,\alpha_z)$ points on the sphere where $B_{abs}$ is an approximately constant scalar and $\hat{\mathbf{n}}$ is the unit normal vector. In the image plots, we see the measured $B_x$, $B_y$, and $B_z$ over each pose $\alpha_y,\alpha_z$ compared to the designed field. There is a small percentage of white pixels representing poses within the workspace that were unachievable by the kinematic processor. The robot scans in a meander, alternating $+z, -z$, and artefacts of this is seen through scan lines in the measured data. Overall, we measure high angular accuracy with a mean error of $2.9^\circ$ and mode error of $2.3^\circ$ and confirm that the robotic arm is able to produce desired field orientations with a high accuracy.
\subsection{Field amplitude control}
For a set vector orientation and magnet mass, some ODMR applications require tuning of the magnetic field amplitude, for instance so that the spin resonance frequency matches a microwave resonator \cite{angerer2018superradiant,putz2014protecting}. The field amplitude can be controlled by tuning the distance between the magnet and the sample position. However, the magnetic field fall-off with $r$ distance is highly non-linear, characterised by the Biot-Savart $1/r^3$ relation. In addition, the robotic arm performs non-linear displacement, requiring dual movement of two rotational joints per linear step of the end effector. 

We observe in Fig. \ref{fig:hall}C that the displacement of the magnet away from the Hall sensor is sufficiently linear to produce a $1/x^3$ response in $B_x$. Because the magnetic field generated by the permanent magnet is large, it can be positioned sufficiently far away from the sensor so that the 10 mm $1/r^3$ trajectory can be subsampled within the 0.5 mm resolution of the robot (lines shown in the top panel) to create a desired response $B(r)$. In the middle panel we observe that we can create a linear field response between 0 and 10 mT through this method. In the bottom panel, we observe the error in this sampling technique is typically lower than 0.1 mT. However, as expected, this error increases for close distances to the sensor, as the available resolution to subsample the $1/r^3$ field diminishes. 

\begin{figure*}[hbtp!]
\centering
\includegraphics[width=\textwidth]{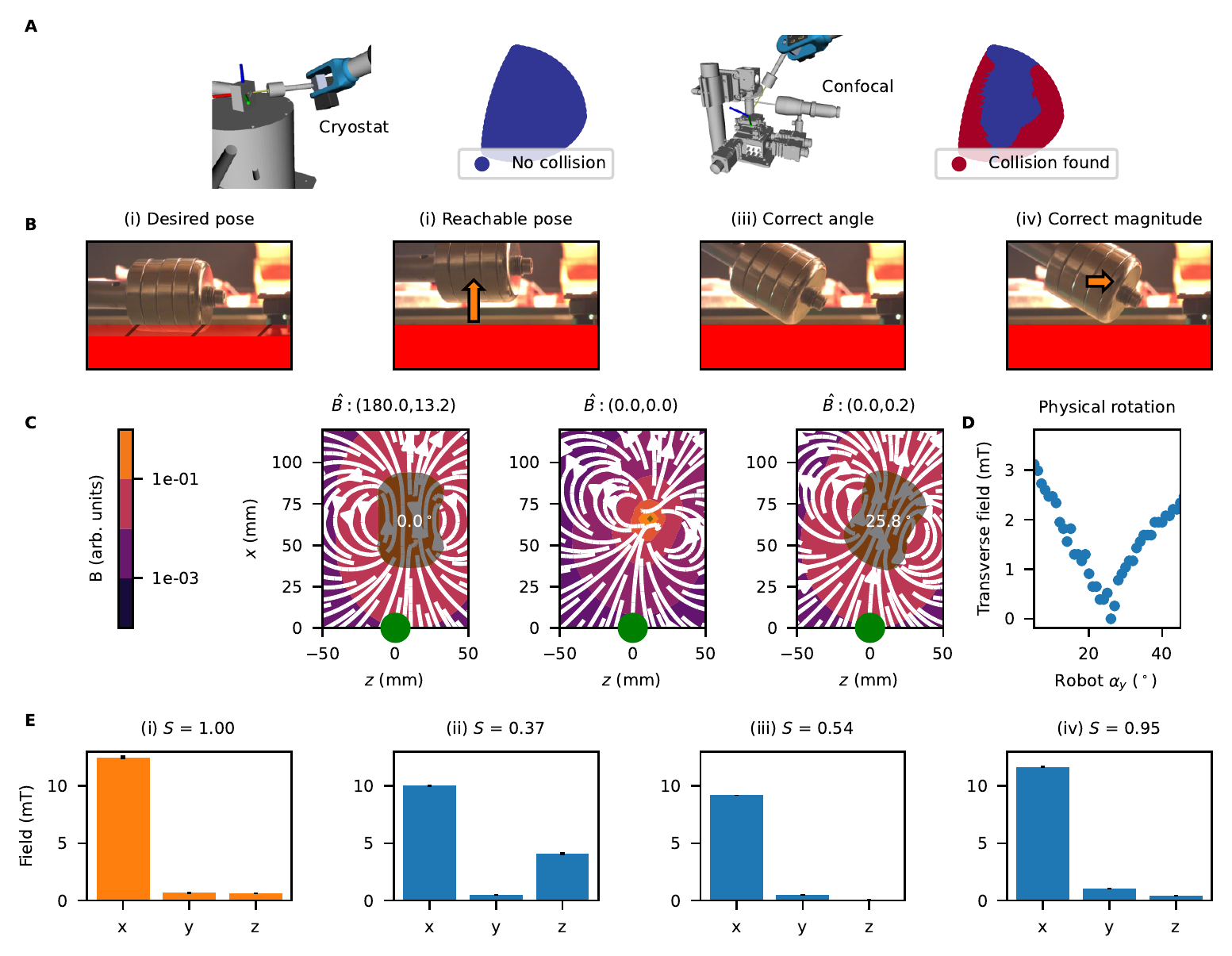}
\caption{\label{fig:collision} \textbf{Motion planning in experimental settings} \textbf{A.} We model two experimental settings, a cryostat with optical access and a confocal microscope. Plotted are the simulated collisions with the robot (in red) and avoided collisions (in blue) over the fixed position varied $(\alpha_y, \alpha_z)$ poses. For the confocal, we observe a limited reachable workspace subset. \textbf{B.} We develop an algorithm to replace these unreachable poses with collision-free poses. The procedure follows: (i) a desired field vector is measured in a forbidden position, (ii) displacement puts the magnet in an allowed position, (iii) angular orientation sets the correct field vector, (iv) a further displacement corrects the field magnitude. \textbf{C.} Using a dipole source to calculate the rotation, we can deterministically rotate the magnet and recover the $\hat{B}$ vector (title bar) at the observer (green dot).  \textbf{D.} The designed rotation matches experimentally in minimising the transverse field at 26 $^\circ$. \textbf{E.} Measuring with the 3D Hall sensor at each stage following panels in B, the final vector well matches the initial vector (quantified by similarity function $S$ defined in the text). }
\end{figure*}

\subsection{Collision-free motion planning}
With operation validated in an unconstrained environment, we move to navigating the robot around complicated lab infrastructure. By evaluating intersections with its environment, the robot is able to compute collisions in the ROS simulation using LBKPiece from the Open Motion Planning Library to traverse a tree of possible trajectories to achieve a given pose goal\cite{sucan2012open}. 

We take two  experimental setups in our lab, a cryostat with an optical window and a scanning stage confocal microscope (see Fig. \ref{fig:collision}A), and add their spatial meshes in simulation to the robot environment.  For these complex geometries, it would not be possible to position 3-axis Helmholtz coils for magnetic field alignment due to the competing requirements for optical access to the sample and the need to move the sample in three dimensions. Single microcoils or a permanent magnet mounted on a stage would have limitations in terms of achievable proximity to the sample. 

With the available kinematics of the 6 DoF robotic arm, the position of the magnet with respect to the sample is far less constrained. In Fig. \ref{fig:collision}A, using LBKPiece, we simulate that a chosen subset of poses can be traversed with access to the top and back side of the cryostat, oriented around the TCP located at the sample mount, generating a $-B_x, +B_y, -B_z$ sphere segment (one-eighth) without collisions. However, we observe that for the scanning stage confocal microscope, only a subset of any sphere segment is achievable. Poses that cannot be accessed without collision are shown with red dots and form a significant part of the subset. From this simulation, it is evident that there would limited success of the robot in the magnetisation-axis-aligned configuration described in Fig. \ref{fig:setup}C. 

\subsection{Designing collision-free field vectors}
An important consideration at this point is that the set of poses in this configuration only make a small subset of the possible joint configurations of the robot, and therefore possible magnetic field vectors. By moving the TCP defined in Fig. \ref{fig:setup}C from the NV centre to the magnet, we give free control on its orientation and position, with access to the fringing fields of the magnetic source. Our hypothesis is that there exists a set of collision-free poses that would produce a full set of magnetic vectors. This idea makes uses of the magnetic inverse problem in field sensing: even if a pose cannot be reached, the desired field can be obtained because there is non-unique mapping between the field and the pose \cite{lima2006}. Our algorithm is laid out along Fig. \ref{fig:hall}B. Firstly, the unreachable set of poses in the constrained environment are found. For each such pose (Fig. \ref{fig:hall}B(i)), the TCP is translated along $x$ to the magnet centre and the magnet is then linearly translated in either $y$ or $z$ to a new reachable pose (Fig. \ref{fig:hall}B(ii)). Next, the new pose is rotated in $\alpha_y$ or $\alpha_z$ to obtain the same vector field orientation as the original pose (Fig. \ref{fig:hall}B(iii)). Finally, the magnet is translated in $x$ to recover the original magnitude (Fig. \ref{fig:hall}B(iv)).

To calculate the vector rotation in Fig. \ref{fig:hall}B(iii), we can approximate the magnet with a dipole, for which the inverse magnetostatic expression is known \cite{lima2006}. For a powerful permanent magnet, the arm can be withdrawn to sufficient distances so that this dipole approximation becomes valid. The orientation of a unit dipole $\vec{m}$ at a vector $\vec{r}$ to create a field at the sensor location $\vec{B}$ is given by:
\begin{align}\label{eq:di}
\vec{m}=\frac{6 \pi}{\mu_0}(\vec{B}\cdot\vec{r})\left|\vec{r}\right|\vec{r}-\frac{4 \pi}{\mu_0}\left|\vec{r}\right|^3\vec{B}
\end{align},
where $\mu_0$ is the vacuum permeability. 

In Fig. \ref{fig:collision}C, we model the $z$ displaced magnet and find the field observed at the sample location (green dot) has a significantly modified orientation (first panel). We can use Eq.\ref{eq:di} to calculate the dipole orientation $\vec{m}$ to find $\vec{B}$ (middle panel). We then rotate the magnet to be coaxial with the calculated dipole orientation and recover the desired field vector $\vec{B}$ with high accuracy, minimising $B_y$ and $B_z$ (last panel). In Fig. \ref{fig:collision}D, we show that in the physical experiment, the off-axial field component is indeed minimised when set to the calculated $26^\circ$ angle, and that this algorithm succeeds within the pose resolution limit of the robot.

As well as correcting orientation, for some applications it is important to maintain the field amplitude. This final step in Fig. \ref{fig:collision}B(iv) is achieved by using the known $1/r^3$ Biot-Savart relation of Fig. \ref{fig:hall}C to scale the amplitude, translating the magnet in $x$. To capture both the orientation and amplitude, we define a Gaussian kernel similarity function between the target field vector $B_1$ and the replacement vector $B_2$, as this is well bounded between 0 (least similar) and 1 (most similar):
\begin{align}
S=\exp{\frac{||B_2-B_1||^2}{2d^2}} 
\end{align}
with $d = 3$. We experimentally implement each step of the algorithm in Fig. \ref{fig:collision}E, and see that the final vector achieves a high similarity to the target vector with $S =0.95$. This can also be seen by comparing the histograms of the measured field components in Fig. \ref{fig:collision}E(i) and Fig. \ref{fig:collision}E(iv). Here, the final amplitude correcting step maintains the desired field orientation. We have evidenced that by using this algorithm, it is possible to systematically replace unreachable poses with reachable poses with the same field vector. The full dexterity offered by the robotic links combined with the inverse problem of magnetostatics make this system a powerful tool for setting arbitrary strength magnetic field vectors in highly constrained environments.   

\begin{figure*}[htbp!]
\centering
\includegraphics[width=\textwidth]{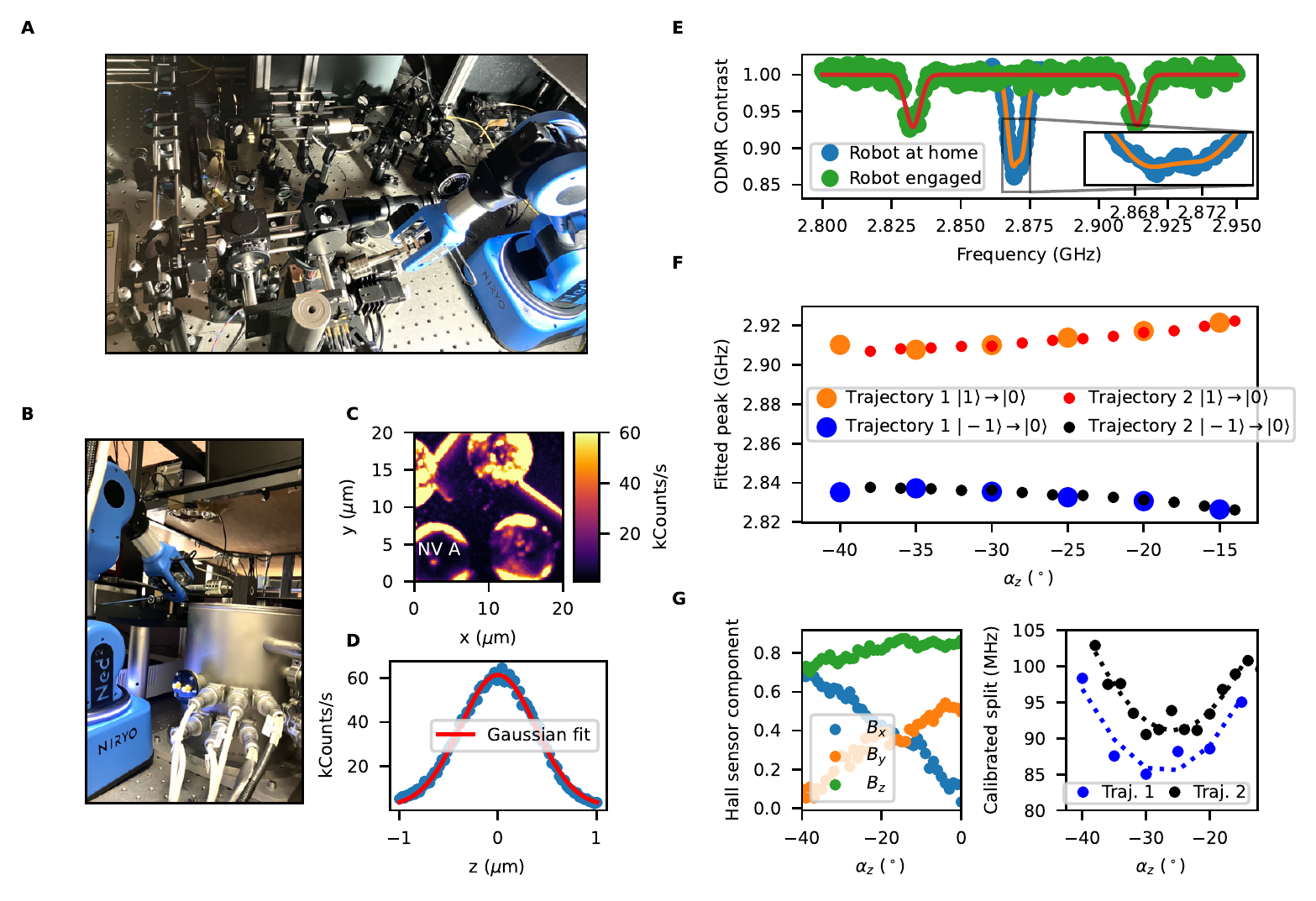}
\caption{\label{fig:odmr} \textbf{Robot-assisted magnetometry}: \textbf{A.} Image of confocal microscope setup showing robotic arm in position. \textbf{B.} Optically accessed cryostat with robotic arm in position \textbf{C}. Confocal image of NV centre located in diamond lens (mounted in setup shown in A) \textbf{D.} Photoluminescence scan in z of above. \textbf{E.} Optically Detected Magnetic Resonance (ODMR) showing zero-field magnetic field splitting of associated spin in blue when robot arm is approximately 10 cm from sensor, and strong 100 MHz splitting in green when robot arm is proximal to the spin sensor. \textbf{F.} Fitted peak resonance for robot trajectory in 5$^\circ$ increments (large circles) and 2$^\circ$ increments (small circles) indicating movement is stable and repeatable. \textbf{G.} Hall sensor data shows the $B_x$, $B_y$ crossover in trajectory. Normalising splitting between resonances by $B$ magnitude reveals $\sigma_z$ dependence in Trajectory 1 (black) and Trajectory 2 (blue), indicative of angular alignment. 
}
\end{figure*}

\subsection{Experimental setting of an NV centre confocal microscope}
Following validation with the Hall sensor, we now move to a full experimental setting in order to evaluate the performance of the robot for aligning a spin based quantum sensor. We see in Fig. \ref{fig:odmr}A and Fig. \ref{fig:odmr}B that this requires navigating a highly complex environment with many sensitive optical and mechanical instruments. 

In the technique described in the previous section, the collision-free positions found in simulation can be downloaded to the physical robot. This requires a fine alignment between the simulation and the real world performance which could be achieved using the approache described in Fig. \ref{fig:hall}A. In a highly constrained environment, we find sensor-driven fine angular alignment is difficult to achieve as the initial calibration trajectories contain poses that cannot be verified as collision-free until this alignment has succeeded. 

A fast and pragmatic approach in an experimental setting is to \emph{teach} the robot a set of collision-free poses. We do this by switching the torque to each motor off momentarily, allowing the joints to move freely. Now, the operator can grasp the magnet, and guide the robotic arm to a desired location within the geometry, avoiding collision. Whilst doing this, the user can monitor the magnetic field produced at the sample location with the Hall sensor. When a desired field is registered, the torque and closed loop feedback can be switched on, locking the magnet at its set position. The corresponding coordinates (either the pose or the joint angles) can be registered along with the measured magnetic field. Through this method, the user can hunt and find locations where, for instance, $z$ is the dominant field component. These poses can be used to gather information to calibrate the source to its experimental surroundings.  We can both compare the taught poses with the simulation to calculate collisions and, with the dipole field model, we can locally modify the taught position to achieve desired magnetic field vectors. 

As a proof-of-principle experiment, we teach the robot a trajectory across the confocal microscope shown in Fig. \ref{fig:odmr}A. We can then demonstrate that traversing this trajectory repeatably affects the NV centre spin based sensor. In Fig. \ref{fig:odmr}C, we use the confocal microscope to locate a collection of milled solid immersion lenses in a polycrystalline diamond sample \cite{hadden2010strongly}, observing a bright NV centre in the central lens (Fig. \ref{fig:odmr}D). With the robot arm at its home position, 10 cm away from the sample, we perform ODMR on this NV centre (see methods for experimental details). As defined in Eq. \ref{eq:Hamil}, we observe a small 3.7030 MHz splitting due to an intrinsic $\Pi =$ 1.8515 MHz, about the central $D = 2.8704$ GHz, shown in Fig \ref{fig:odmr}E. In moving the TCP near to the NV centre site, with the robot end effector in the vicinity of the experimental setup, we observe a large 100 MHz splitting in the ODMR spectra (green plot) from the presence of the permanent magnet. Typically, the NV centre must be optically aligned to the photon detector within 500 nm using a piezo stage and the ability to engage and disengage the robot whilst performing ODMR indicates it is suitable to use in this highly sensitive experiment. 

\subsection{Robotic vectorial field alignment of a single spin sensor}
In Fig. \ref{fig:odmr}F, we repeat the ODMR for a set of robot poses along a collision-free trajectory with $\alpha_y$ = 20 $^\circ$ at 5 $^\circ$ increments in $\alpha_z$ (Trajectory 1). Each data point was averaged over five minutes. The peak resonances show a smooth increase in splitting from $D$ as the magnet is moved in front of the objective. The same path is traversed at 2 $^\circ$ increments in $\alpha_z$ (Trajectory 2) and resonance data shows the same splitting trend, indicating that the trajectory-induced splitting is reproducible. 
The fitted resonances can be used to extract the polar angle between the NV axis and the magnetic field of interrogation, from the zero-field parameters $D$ and $\Pi$ (see Methods for details) \cite{balasubramanian2008nanoscale}. This gives a value of $61.0\pm 0.3^\circ$  for Trajectory 1 and
$61.8\pm0.2^\circ$ for Trajectory 2 \cite{balasubramanian2008nanoscale}. This small polar angle change over the 25 $^\circ$ range in $\alpha_z$ indicates that the NV centre is near aligned with the world coordinate z-axis. 

In Fig. \ref{fig:odmr}G, replacing the diamond sample with the Hall sensor, and focussing the objective at the centre of the sensor area, we can map the field components generated by the trajectory. We see that the $B_x$ and $B_y$ components cross, as expected for varying $\alpha_z$. We find the Hall field amplitude $|B|$ in excellent agreement with the ODMR sensed field amplitude (see Supplementary Fig. 3). For the trajectory chosen, the robot TCP is not yet fully aligned with the NV or Hall sensor area. We see that in contrast to Fig. \ref{fig:hall}B, the trajectory does not conserve $|B|$, with a large increasing $B_z$ component, and this is the main contribution to the increased splitting between the two resonances in Fig. \ref{fig:odmr}F.

Using the amplitude data, we can normalise each splitting to isolate the field component ratios (see methods). In Fig. \ref{fig:odmr}G, this normalisation results in the appearance of a dip from both trajectories, with the phase and contrast of the dip giving an estimation of the NV's orientation. The low contrast, non-zero minima in $\alpha_z$ indicates the NV centre is near aligned with the world coordinate z-axis as previously discussed. We find that both datasets can be well fitted with the characteristic equation \cite{fukushige2020identification} to find $\alpha_z = 64.1 \pm 0.4^\circ, \alpha_y=97.6 \pm 0.7^\circ$ for Trajectory 1 and $\alpha_z= 62.9 \pm 0.8^\circ$, $\alpha_y=97.9 \pm 1.4^\circ$ for Trajectory 2.
In this, we have shown that vectorial field alignment can be achieved with robotic trajectories of as little as six steps. This is especially useful for ODMR where collecting each spectra takes on the order of minutes. 

\section{Discussion}
Our results show that an industrially designed robotic arm can be adapted to operate around sensitive optomechanical samples and setups. The presented modality produces stable and controllable magnetic fields that are capable of manipulating and aligning a single solid state quantum spin sensor. This an important step in the use of robotics to replace axial stages and bulky field coils for experimental physics and in developing quantum technologies, where we have evidenced the benefit of the innate flexibility and configurability of robotics arms in highly constrained environments.  

The next step for this work is to generate on-demand magnetic fields using a sophisticated algorithm that maps the traversable space given geometrical parameters, making use of the collision-free techniques described. With this, a set of control points can be found, considering application-specific criteria such as the field magnitude, linearisation, or the time taken to move between points.

Robotics, unlike solenoid coils, produce minimal local heat. This makes them suited for sensitive samples and algorithms could be designed for tracking quantum sensors in motion under cell uptake, a difficult task where the spin sensor orientation changes over time \cite{le2013optical, rendler2017optical}. For further flexibility, the cylindrical magnet could be replaced with a rectangular magnet fixed perpendicular to its magnetic axis, with the unused roll degree of freedom in the robotic wrist providing rapid field orientation. 

Beyond an off-the-shelf design, an application-specific robot could further maximise efficiency, precision, and control. This could have a larger payload whilst having a smaller form factor, for instance. We can extend this to the use of multiple robots to generate gradient magnetic fields. As well as a range of solid-state sensors, the alignment of atoms and ions in cold and vacuum environments can be explored with these form factors.

In addition, the robot-driven orientation presented can be extended to aligning quantum objects with a range of parameters including electric and lights fields. Here, the end effector would be an electrode, or in optics, a laser or mirror surface. Following this proof-of-principle work, the adaptability of robots in combination with sophisticated software could provide ruggedness for alignment in demanding real-world environments where quantum technologies are emerging such as point to point Quantum Key Distribution (QKD) \cite{sidhu2021advances} and quantum rangefinding \cite{frick2020quantum}.

\section{Methods}
\subsection{Magnetic field modelling}
Magnetic field calculations in this work are performed using the closed form expressions presented in the Magpylib package\cite{magpylib2020}. The hollow cylindrical magnet is modelled by subtracting an inner cylindrical magnetic source of opposite magnetisation from the outer cylindrical magnet source. 

\subsection{Robotic modelling}
The Niryo NED 2 robot geometry is specified in the Unified Robot Description Format (URDF). Here, the end-effector geometry file specified in the URDF is replaced with a geometry of the magnet tool. For the collision-free motion path finding, the experimental setups are modelled in FreeCad and replace the geometry file of the robot base. The robot is simulated in a ROS environment, and controlled using the python wrapper PyNiryo2.

\subsection{Experimental setup}
The magnetic field measurements in this work is made using the Infineon TLE493D-P2B6MS2GO 3D Magnetic Sensor fitted on a compact platform mount or in the described confocal microscope. 

For the ODMR measurements, the NV centre is excited by a CW 532 nm laser (gem 532; Laser Quantum). A confocal microscope is used to image the collected count rate. Using a 0.9 NA microscope objective, the excitation beam is highly focused on the sample producing a nearly diffraction-limited spot (< 1 \textmu m diameter). The NV centre PL is collected through the same lens and separated from the excitation path by the use of a dichroic mirror before detection by single photon avalanche diodes (SPADs)(SPCM-AQRH-12-FC; PerkinElmer). By scanning the position of the sample, a map of detected count rate is generated by which the position of the NV centre and its maximum count rate can be found. ODMR was performed under CW excitation using a Rohde and Schwarz SMB100A microwave source driving a custom loop antenna PCB on which the sample is mounted.

\subsection{Spin sensor modelling}
The spin transitions presented in Fig. \ref{fig:setup}B are calculated solving the Hamiltonian eigenstates in QuTiP \cite{johansson2012qutip}. Other calculations use the NV spin energies characteristic polynomial presented in Balasubramanian \emph{et al.}, whereas the polar angle $\theta$ between the field and the NV centre is found using the solution given in that work \cite{balasubramanian2008nanoscale}. The polynomial can be solved and least-squared fitted to the experimental data to find the NV orientation $(\alpha_y^{\mathrm{NV}},\alpha_z^\mathrm{NV})$: \cite{fukushige2020identification}, 
\begin{align}
    \begin{split}
x^3-\left(\frac{D^2}{3}+\Pi^2+\beta^2\right) x-\frac{\beta^2}{2}D \cos 2 \gamma&-\frac{D}{6}\left(4 \Pi^2+\beta^2\right)\\&+\frac{2 D^3}{27}=0.
    \end{split}
\end{align}
where $\gamma = \arccos \left(\left|\cos \left(\alpha_z^{\mathrm{B}}-\alpha_z^{\mathrm{NV}}\right) \cos (\alpha_y^{\mathrm{B}}-\alpha_y^{\mathrm{NV}})\right|\right)$ with a known $\alpha_y^{\mathrm{B}}$ and  $\alpha_z^{\mathrm{B}}$ set by the robot, $D$ and $\Pi$ fitted from the zero-field data, and $\beta = \gamma_e |B|$ where $\gamma_e$ is the gyromagnetic ratio and $|B|$ is the external magnetic field amplitude. In the trajectories presented, and the separation between resonances $\nu(i)$, $|B|$ is not conserved. For this fit, we must obtain a constant $|B|$ so first normalise using the field magnitude data. For this, we sub-sample the higher resolution Hall data to reduce noise (see Supplementary Fig. 3) and obtain normalised splittings $\nu_n(i)$ for each measurement point $i$ where:
\begin{align}
\nu_n(i) = \frac{\nu(i)}{|B_\text{Hall}(i)|}\max |B_\text{Hall}|
\end{align}
and leave the non-physical $B$ in the characteristic equation as the third free parameter in the fit to this data. 

\section*{Acknowledgements}
We thank Jorge Monroy-Ruz for building the NV centre confocal microscope used in this experiment. We thank him, Hao-Cheng Weng, Wyatt Vine and John G. Rarity for useful discussions. We acknowledge funding support from the  Engineering and Physical Sciences Research Council (EPSRC)  grant  QC:SCALE  EP/W006685/1.

\bibliographystyle{unsrt}
\bibliography{sample}

\end{document}